# Transforming Critical Spreadsheets into Web Applications at Zurich Financial

Sebastian Dewhurst (EASA)
Davenport House, 39 Evenlode Drive, Wallingford, OX10 7NZ, UK.
seb.dewhurst@easasoftware.com

**ABSTRACT**

*In the insurance industry, spreadsheets have emerged as an invaluable tool to for product pricing, because it is relatively straightforward to create and maintain complex pricing models using Excel®. In fact, Excel is often preferred to "hard-code" whenever there are frequent changes to the calculations and business logic which under-pin the pricing of an insurance product.*

*However, problems arise as soon as spreadsheets are deployed to end-users: version control, security of intellectual property, and ensuring correct usage are obvious issues; frequently, integration with other systems is also a requirement.*

*Zurich Financial Services Group is a leading financial services provider; several possible solutions to these problems have been evaluated, and EASA has been selected as the preferred technology. Other spreadsheet collaboration approaches which were considered include Excel Services, and/or custom-built software; however, EASA has provided clear benefits over these strategies.*

## 1. INTRODUCTION

In the financial sector, the proliferation of un-centralized, end-user applications such as spreadsheets is driven by many factors. One factor is the degree of specialisation within the industry [Sentance, 2006]. Specialisation of knowledge (for example, that of actuaries), combined with the need to update the logic embedded in end-user applications on an almost real-time basis, conspire to make it almost impossible to deliver solutions using conventional IT technologies.

As a result, many key processes at Zurich are under-pinned by spreadsheets; they are easy to create, familiar to users, and flexible. However, there has historically been a high cost associated with deploying these spreadsheet-based processes, while keeping the risk of mistakes acceptably low and maintaining version control and correct usage.

Insurance rating is a specific example of such a process. It is a critical one in that the calculations, methodologies and business logic embedded in rating models can contribute directly to the company's bottom line – an error would have a profound financial effect. Not only that, but these algorithms and methods, usually developed by actuaries, represent key intellectual property; it is vital that they are applied correctly and protected from unauthorized distribution.

Solutions that involve simply "locking-down" spreadsheets have been investigated, but this approach is fraught with limitations. At the other extreme, the wholesale replacement of operational spreadsheets with custom applications has been considered. However, while there is occasionally a case for eliminating spreadsheets [Powell, 2009], this approach is usually expensive and time consuming.

If we accept, then, that spreadsheets will always fill a void between the business needs of today and formal installed systems [Baxter, 2007], and that the complete elimination of spreadsheets is unlikely, what can be done to ensure that where spreadsheets *are* used, their usage is appropriately managed? Weiss [2006] outlines five key targets for managing spreadsheet-based processes:

1. Financial professionals are comfortable with spreadsheets; don't try to change their working environment. Instead, look for a solution in which the user continues to work almost entirely in Excel, but add appropriate tracking and management functions (in the background, where possible).

2. Ensure secure user access; users should only have access to mission-critical spreadsheets based on their permissions and privileges.

3. Provide change and version control; it should be absolutely impossible for an end-user to use an out-of-date spreadsheet.

4. Automate the review-and-approval process; eliminate error-prone review processes by providing a secure repository and a trackable process by which an authorized person can approve a particular version of an updated spreadsheet for the end-users.

5. Retain essential spreadsheets for your records; if business decisions have been made with a particular version of a spreadsheet, then in certain situations it may become necessary to "roll-back" to that version to re-create a specific report or calculation.

One further point: most of our critical spreadsheets contain macros; thus, they cannot be deployed with Excel Services without a cost-prohibitive amount of re-work.

## 2. PROJECT OVERVIEW

The main objective is to deliver a single set of integrated tools that support our Sales and Underwriting functions, and to provide a foundation upon which Zurich's Global Corporate achieves its Underwriting Target Operating Model (UW TOM) objectives of becoming a globally integrated organization. Particular focus is on the delivery of new, consistent, pricing tools.

Within this context, the re-deployment of several existing spreadsheets as robust, secure, enterprise-class web applications is considered a key component of the UW TOM. Without this, Global Corporate would not be able to achieve the benefits of a globally integrated business.

### 2.1 Key drivers

Key drivers include:

- The need to improve cost efficiency and effectiveness of existing and future sales and underwriting processes;

- The need to improve quality and availability of decision-making, operational and management information;

- The need to provide our customers with consistent pricing, under-pinned by methodologies that are "certified";

- The need to provide consistent execution of the "Zurich Way of Underwriting";

- The need to provide the capability for Global Corporate and ultimately Group Global Customer clearance;

- The need to provide a consistent Customer and Whole account view of the Global Corporate portfolio.

### 2.2    Preferred Solution

Zurich selected EASA (Enterprise Accessible Software Applications), a commercially available tool for building Web-based applications which leverage existing assets such as spreadsheets, databases, and legacy applications.

EASA's spreadsheet management solution allows us to secure a master version of a given spreadsheet on a server. Authorized users may then access it only via a custom web application created with EASA's codeless application builder, allowing a more natural work-flow. Ultimately, users will access these tools via a proprietary portal which calls the custom EASA applications via web-services.

- The custom web application is so intuitive that training is no longer required (see test-case, Figure 1);

- Users only see what they need to see; they no longer access the spreadsheet directly, and are not able to make unauthorized changes with it;

- If a change is required, it is straightforward for an authorized person (typically an actuary) to make a change to the master version of the spreadsheet and re-upload it. It is immediately published to all users, and so version control is assured;

- Integration with other corporate systems can be achieved because EASA supports web-services;

- Even spreadsheets which contain add-ins or macros can be deployed, because EASA runs Excel natively on a server; this is in contrast to technologies which simply translate the original spreadsheet into a database application.

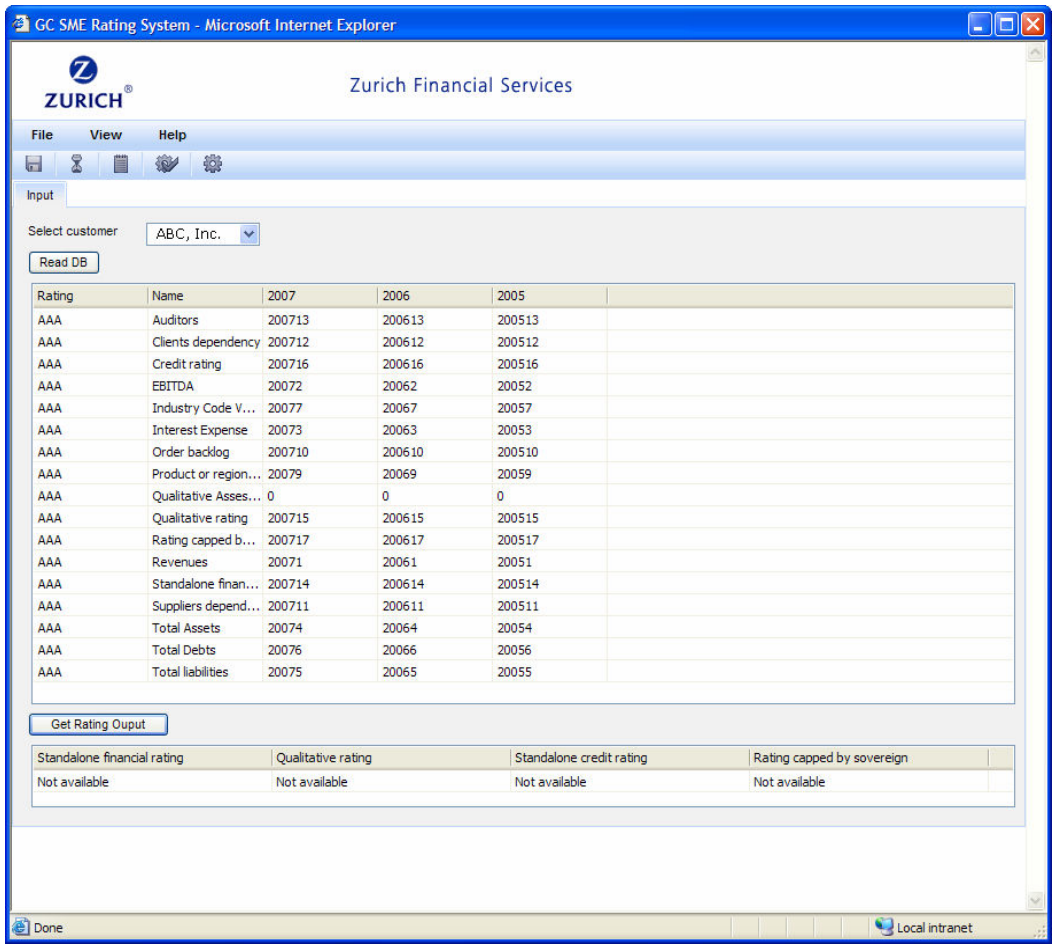

**Figure 1. A test-case was created to demonstrate the ability to rapidly create function-specific web-applications based on existing spreadsheets.**

### 3. OVERVIEW OF EASA

EASA has been in use since 2002, by industries as diverse as health-care, communications, energy, financial services, and manufacturing. Specific uses have included the provision of custom interfaces to multiple existing applications, giving users simplified access to key software tools and data [Kornfein & Rajiv, 2008]. Another common purpose is the modernization of legacy software, which can be "wrapped" by EASA and transformed into modern, web-enabled applications, accessible to any desktop, laptop, or mobile device in the enterprise [Casanova, 2003]. EASA's architecture is shown in Figure 2.

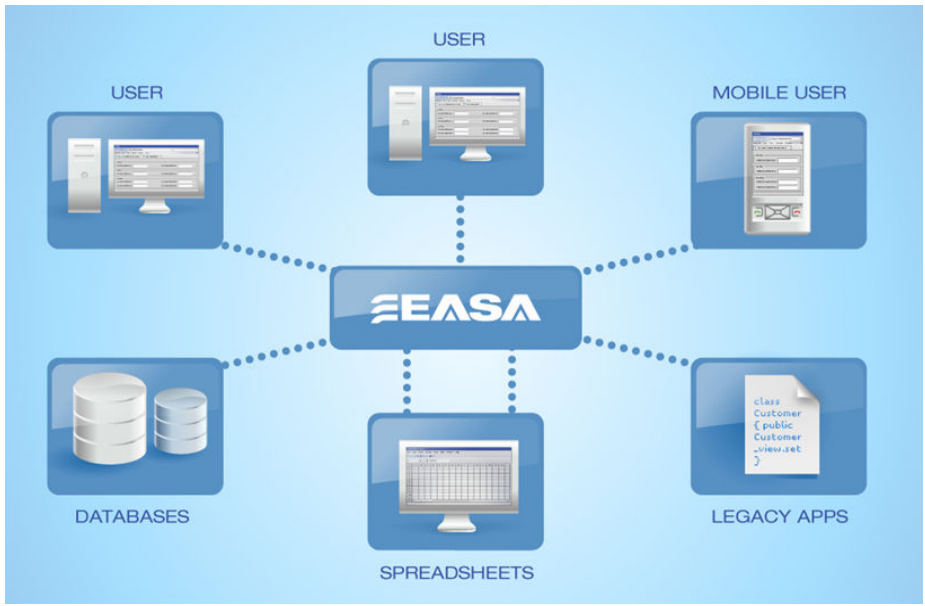

**Figure 2. EASA architecture**

EASA enables the creation of simple web based applications (or EASAPs) that link to one or more key spreadsheets, databases, and other existing software. This not only eliminates the need to distribute key spreadsheets, but also ensures that they are used *precisely* the way their authors intended, reducing user-errors. A typical EASAP is shown in Figure 3.

**Figure 3. A typical EASA application, or "EASAP", which connects to a central spreadsheet**

A side benefit includes the ability to link spreadsheets with other enterprise software, and to provide user-specific views – only relevant information is exposed to users. Finally, it is possible to establish a record of who did what, with which spreadsheet, and when.

The results pages provide an option to share reports on completed work with other users; their content is defined by the author, and can contain text, tables, charts, images or even animations generated by the application. Figure 4 shows a typical example.

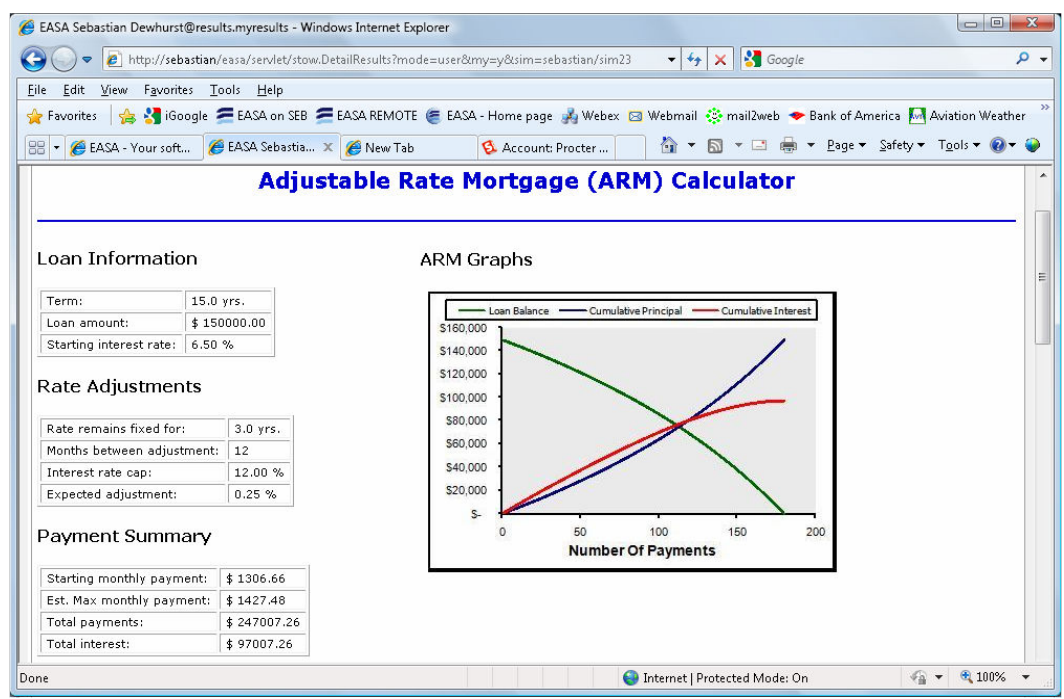

**Figure 4. Example Report Page from EASA**

A custom EASAP (EASA Application) can drive multiple underlying software tools installed on existing systems throughout the enterprise. The underlying software may be anything from complex "expert only" applications, to legacy systems, modern databases, and spreadsheets.

Each EASAP is built for a specific need within the organization. EASAPs are available over the intranet to authorized users throughout the enterprise, providing secure, simplified access to the company's processes, best practices, expertise and software assets.

EASA's codeless application builder is used to create a user interface, to link the interface to underlying software and databases, and if required, to create custom reports. This eliminates complicated and time-consuming coding of custom applications with tools such as C++, VB, and Java and their associated Integrated Development Environments, which might take weeks, months, or even longer. By comparison, EASA allows new custom applications to be created, tested, and deployed in as little as a few hours.

EASA is a client-server environment, in which the information needed to run a particular software application is locked into an EASAP by the author. An EASAP can drive any batch-capable or COM-compliant software (and has dedicated wizards for linking to Excel), and also allows interaction with JDBC compliant databases. EASA contains the necessary structures to run applications on different computers (where they are resident), to control queuing and user access, and to serve up web-pages to the users.

The authoring tools are a particularly powerful part of the EASA system, and a typical EASAP can be created in 2 to 8 hours. EASAP Builder provides a tree structure for building the EASAP and linking it to a centrally maintained spreadsheet, databases and/or other software assets. No knowledge of programming is needed. Figure 5 shows an example of an author's tree for a typical EASAP.

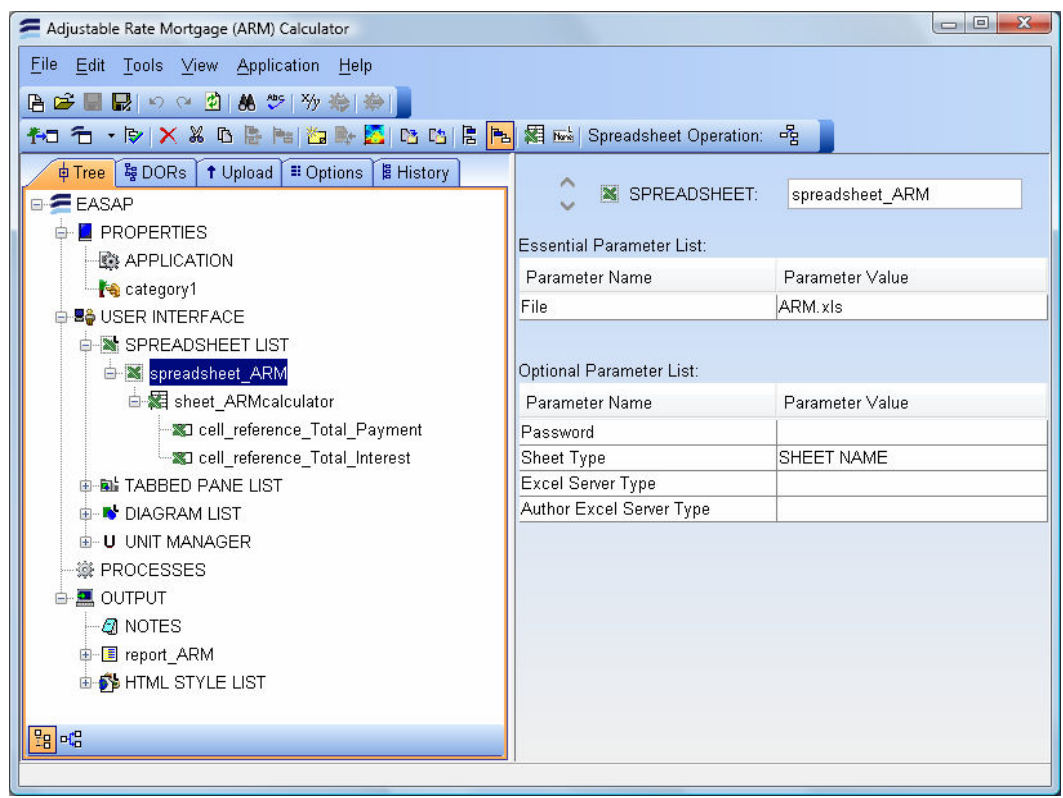

**Figure 5. EASAP Builder – typical screen**

EASAP Builder allows an author to create:
- user-interface components such as tabbed panes, choice lists, radio buttons, and input fields;
- components that validate data typed into the input fields;
- components that perform calculations from this by passing it to a spreadsheet (or other programs);
- user-interface components that display results.

EASA also provides a version control system for the EASAPS – allowing them to be developed "off line" by an Author, then, when ready, to publish them on the Intranet. Any subsequent versions are then automatically given revision numbers. Previous versions of EASAPS can be restored by the system administrator.

## 4. LESSONS LEARNED

(The point of contact at Zurich is Nemanja Kostic. He may be contacted at nemanja.kostic@zurich.com for more details).

Some lessons learned from this project include:

- The learning curve was not steep; after initial two-day training courses for authors, the overhead in terms of publishing Excel spreadsheets as EASAPs is from a few hours to one or two days (per application). Maintenance overhead is extremely low. As a result, it has not been difficult to change the behavior of both authors and users.
- In fact, the behavioural changes required were minimal. Users can no longer use their own unapproved versions; instead, spreadsheets from actuaries are sent to the EASA administrator for deployment, which usually takes only a few minutes; in other words, there is no delay in getting approved, current tools into the hands of users (compared to the old approach of circulating raw spreadsheets). The process is very strict and defined. While this represented a change for the underwriters, who were accustomed to using their local versions, the new web based applications rapidly become popular, once the process has been clearly communicated and demonstrated.
- Discipline with respect to the underlying spreadsheets is important. Executing Macros, VBA, or menu commands in an underlying spreadsheet are fine. Indeed this is a differentiator for the EASA technology, and most of Zurich's cases, the user executes macros defined in spreadsheets via buttons on new interface. However, error-trapping does need to be acceptable (e.g. write errors to a cell and thence to the application, so that end-users are not left without feed-back in the event of a failure). This required removal of pop-ups from any macros. It was a minor change to our existing spreadsheets.
- Similarly, named ranges are far more reliable than specific cell references, in terms of future maintenance of these Excel-based applications. Input fields and ranges are named so that the link between EASA and a spreadsheet is name-based, rather then position-based. Thus, input/output fields may be moved around the spreadsheet without rewiring the whole EASA application.
- Frequently, applications may be updated via a change purely to the underlying spreadsheet, versus a change to the actual application. This is obviously tempting, as it is the work of a few minutes (in addition to the actual time taken to update the spreadsheet). However, sometimes it is more efficient in the long-run to update the actual application, re-test, and re-publish. This takes longer (perhaps an hour), but is still faster than editing hard code.
- Moving a desktop application to an enterprise environment opens up many advantages, such as leveraging more robust database technologies where appropriate. However, care must be taken when enabling interaction between Excel and other applications. A specific example is the date format used in Excel, which does not necessarily interact well with databases. EASA does in fact provide capabilities to translate date formats, but it is nonetheless a general issue to keep in mind
- Robustness of Excel was a concern at the beginning of this project – there is a good reason why MS has developed Excel Services. However, in practice, EASA manages this very effectively by closing down Excel processes after use (thus lowering the frequency of "crashes" to almost zero, and also eliminating memory leakage issues). Failover of a specific Excel instance is also contained by the fact that one EASA Server

"brokers" Excel processes to many (perhaps tens of) Excel Servers. In practice, we believe that it's better to have a single point of failure with a single point of control, than to have a "fail-safe" approach without any control at all.

- No practical limit was found in terms of size of spreadsheets – spreadsheets up to 30MB and 300,000 populated cells have been successfully deployed as web-applications with this approach.
- No practical limit of performance has been determined. In fact, because the spreadsheet itself is not rendered for the client, performance is sometimes superior when accessing a spreadsheet via EASA compared to doing so via Excel on a desktop, especially when complex macros are executed.
- There are (currently) some Excel features not supported with this technology, such as references to other spreadsheets. The latter is an inherently difficult issue, as there is no easy way to "abstract out" such a reference. Nonetheless, EASA supports the vast majority of Excel features, and in fact was selected over Excel Services for this reason.
- In our case, the user reaction has been very positive. We are now migrating more spreadsheets into EASA.
- Total cost of ownership of this approach is acceptably low, at ~$1k per end-user.

## 5. CONCLUSION

The existing spreadsheets within Zurich can now be leveraged without any significant re-work. An approach has been established which enables control over the use of the most critical spreadsheets, while allowing for integration of these spreadsheets with other corporate software systems (Figure 6). In the future the system may be extended to provide access to critical spreadsheets for mobile users.

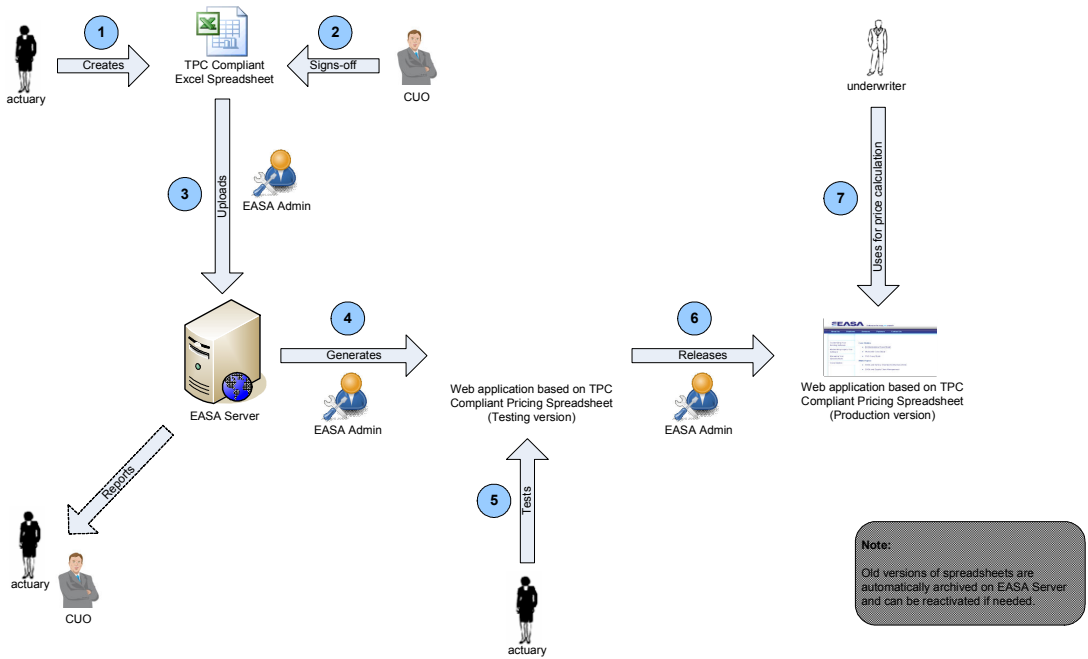

**Figure 6. Zurich Financial Services has delivered a single set of integrated tools that support our Sales and Underwriting functions, while leveraging the business intelligence already contained in many existing spreadsheets.**